\documentclass[preprintnumbers,amsmath,amssymbm,prd]{revtex4}
\usepackage{epsfig}
\usepackage{graphicx}

\begin{document}
\title{On the branching of the quasinormal resonances of near-extremal Kerr black holes}
\author{Shahar Hod}
\address{The Ruppin Academic Center, Emeq Hefer 40250, Israel}
\address{ }
\address{The Hadassah Institute, Jerusalem 91010, Israel}
\date{\today}

\begin{abstract}
\ \ \ It has recently been shown by Yang. et. al. [Phys. Rev. D {\bf
87}, 041502(R) (2013)] that rotating Kerr black holes are
characterized by {\it two} distinct sets of quasinormal resonances.
These two families of quasinormal resonances display qualitatively
different asymptotic behaviors in the extremal ($a/M\to 1$)
black-hole limit: The zero-damping modes (ZDMs) are characterized by
relaxation times which tend to infinity in the extremal black-hole
limit ($\Im\omega\to 0$ as $a/M\to 1$), whereas the damped modes
(DMs) are characterized by non-zero damping rates ($\Im\omega\to$
finite-values as $a/M\to 1$).
\newline
In this paper we refute the claim made by Yang et. al. that
co-rotating DMs of near-extremal black holes are restricted to the
limited range $0\leq \mu\lesssim\mu_{\text{c}}\approx 0.74$, where
$\mu\equiv m/l$ is the dimensionless ratio between the azimuthal
harmonic index $m$ and the spheroidal harmonic index $l$ of the
perturbation mode. In particular, we use an analytical formula
originally derived by Detweiler in order to prove the existence of
DMs (damped quasinormal resonances which are characterized by {\it
finite} $\Im\omega$ values in the $a/M\to 1$ limit) of near-extremal
black holes in the $\mu>\mu_{\text{c}}$ regime, the regime which was
claimed by Yang et. al. not to contain damped modes. We show that
these co-rotating DMs (in the regime $\mu>\mu_{\text{c}}$) are
expected to characterize the resonance spectra of rapidly-rotating
(near-extremal) black holes with $a/M\gtrsim 1-10^{-9}$.
\end{abstract}
\bigskip
\maketitle

\section{Introduction}

Perturbed black holes display a unique pattern of damped
oscillations, known as quasinormal resonances, which characterize
the relaxation phase of the black-hole spacetime. The spectrum of
quasinormal resonances reflects the physical parameters (such as
mass, charge, and angular momentum) of the black-hole spacetime.

The complex quasinormal resonances correspond to perturbation fields
which propagate in the black-hole spacetime with the physically
motivated boundary conditions of purely outgoing waves at spatial
infinity and purely ingoing waves crossing the black-hole horizon
\cite{Detw}. These boundary conditions single out a discrete set,
$\{\omega^{\text{QNM}}(n;m,l)\}_{n=0}^{n=\infty}$, of complex
black-hole resonances for each perturbation mode (here $m$ and $l$
are the azimuthal harmonic index and the spheroidal harmonic index
of the wave field, respectively).

In a very interesting paper, Yang et. al. \cite{Yang} have recently
studied numerically the quasinormal spectrum of near-extremal
(rapidly-rotating) Kerr black holes. The authors of \cite{Yang} have
reached the remarkable conclusion that these rapidly-rotating black
holes are characterized by {\it two} qualitatively distinct sets of
quasinormal resonances:
\begin{itemize}
\item {Zero-damping modes (ZDMs), which are characterized by the
asymptotic  property \cite{Hod1}
\begin{equation}\label{Eq1}
\Im\omega^{\text{ZDM}}(\tau\to 0)\to 0\  ,
\end{equation}}
and \item {Damped modes (DMs), which are characterized by the
asymptotic property
\begin{equation}\label{Eq2}
\Im\omega^{\text{DM}}(\tau\to 0)\to {\text{finite values}}\ .
\end{equation}}
\end{itemize}
Here
\begin{equation}\label{Eq3}
\tau\equiv {{r_+-r_-}\over{r_+}}
\end{equation}
is the dimensionless Bekenstein-Hawking temperature of the black
hole \cite{Notetem}, where $r_{\pm}\equiv M\pm (M^2-a^2)^{1/2}$ are
the black-hole (event and inner) horizons. This dimensionless
temperature approaches zero in the extremal $a\to M$ ($r_-\to r_+$)
limit of rapidly-rotating black holes.

\section {The erroneous claim made in \cite{Yang} and Detweiler's damped resonances}

It has been asserted in Ref. \cite{Yang} that the ZDMs (\ref{Eq1})
exist for all co-rotating modes ($m\geq 0$) \cite{Hod1}, whereas the
DMs (\ref{Eq2}) exist for counter-rotating modes ($m<0$) and for
co-rotating modes in the {\it limited} range
\begin{equation}\label{Eq4}
0\leq \mu\lesssim\mu_{\text{c}}\  .
\end{equation}
Here
\begin{equation}\label{Eq5}
\mu\equiv {{m}\over{l}}\
\end{equation}
is the dimensionless ratio between the azimuthal harmonic index $m$
and the spheroidal harmonic index $l$ of the perturbation mode. The
critical ratio, $\mu_{\text{c}}$, is given by
$\mu_{\text{c}}=\sqrt{{{15-\sqrt{193}}\over{2}}}\simeq 0.74$ in the
eikonal limit \cite{Yang,Hod2}. This critical value of the
dimensionless ratio $\mu$ marks the boundary between perturbations
modes (those with $\mu<\mu_{\text{c}}$) which are characterized by
{\it imaginary} values of the angular-eigenvalue $\delta$
\cite{Delta,Teuk} and perturbations modes (those with
$\mu>\mu_{\text{c}}$) which are characterized by {\it real} values
of the angular-eigenvalue $\delta$ \cite{Delta,Teuk}.

In this Comment we would like to point out that the assertion made
in Ref. \cite{Yang}, according to which co-rotating DMs exist {\it
only} in the limited range $0\leq \mu\lesssim\mu_{\text{c}}$ [see
Eqs. (\ref{Eq2}) and (\ref{Eq4})], is actually erroneous. In
particular, we shall show that co-rotating DMs of near-extremal
black holes [see Eq. (\ref{Eq10}) below] actually exist in the {\it
entire} range
\begin{equation}\label{Eq6}
0\leq \mu\leq 1\  .
\end{equation}
In fact, Detweiler \cite{Det} has obtained an analytic expression
for co-rotating DMs of near-extremal black holes which is valid in
the regime $\mu>\mu_{\text{c}}$ \cite{Notemu}:
\begin{equation}\label{Eq7}
\varpi_n\equiv
M(\omega_n-m\Omega_{\text{H}})=-{{e^{\theta/2\delta}}\over{4m}}(\cos\phi+i\sin\phi)\times
e^{-\pi n/\delta}\ ,
\end{equation}
where $\Omega_{\text{H}}\equiv a/2Mr_+$ is the angular-velocity of
the black-hole horizon, and the integer $n$ is the resonance
parameter of the mode. Here we have used the definitions \cite{Det}
\begin{eqnarray}\label{Eq8}
re^{i\theta}\equiv\Big[{{\Gamma(2i\delta)\over\Gamma(-2i\delta)}}\Big]^2
{{\Gamma(1/2+s-im-i\delta)\Gamma(1/2-s-im-i\delta)}\over
{\Gamma(1/2+s-im+i\delta)\Gamma(1/2-s-im+i\delta)}}\ \ \ ; \ \ \
\phi\equiv -{{1}\over{2\delta}}\ln r\  .
\end{eqnarray}
It is worth emphasizing again that the expression (\ref{Eq7}),
originally derived in \cite{Det}, describes DMs in the
$\mu>\mu_{\text{c}}$ ($\delta^2>0$) regime, the regime which was
claimed in \cite{Yang} not to contain DMs.

\section{The source of the erroneous claim made in \cite{Yang}}

It is important to understand the reason for the failure of Yang et.
al. \cite{Yang} to observe the DMs (\ref{Eq7}) of \cite{Det} in the
regime $\mu>\mu_{\text{c}}$ \cite{Notefal}. In order to understand
the null result of \cite{Yang} in finding numerically the DMs
(\ref{Eq7}), one should examine the regime of validity of the
analyzes presented in \cite{Teuk} and \cite{Det}.

A careful check of these analyzes reveals that the expression
(\ref{Eq7}) for the black-hole DMs \cite{Det} is valid in the regime
\begin{equation}\label{Eq9}
\tau\ll |\varpi|\ll x\ll 1\ ,
\end{equation}
where the dimensionless coordinate $x\equiv (r-r_+)/r_+$ belongs to
an {\it overlapping} region in which two different expressions for
the radial Teukolsky wave function (hypergeometric and confluent
hypergeometric functions) can be matched, see \cite{Teuk,Det} for
details. Taking cognizance of the inequalities in (\ref{Eq9}), one
realizes that the expression (\ref{Eq7}) for co-rotating DMs with
$\mu>\mu_{\text{c}}$ is only valid in the regime of near-extremal
(rapidly-rotating) black holes.

In particular, since each inequality sign in (\ref{Eq9}) roughly
corresponds to an order-of-magnitude difference between two
variables (that is, $\tau/\varpi\lesssim 10^{-1}$, $\varpi/x\lesssim
10^{-1}$, and $x\lesssim 10^{-1}$), the expression (\ref{Eq7}) for
the black-hole DMs \cite{Det} is not expected to be valid outside
the regime \cite{Notesaf}
\begin{equation}\label{Eq10}
\tau\lesssim10^{-4}\  .
\end{equation}
The inequality (\ref{Eq10}) corresponds to rapidly-rotating black
holes with [see Eq. (\ref{Eq3})]
\begin{equation}\label{Eq11}
{{a}\over{M}}\gtrsim1-10^{-9}\  .
\end{equation}


It is worth noting that the numerical analysis presented in
\cite{Yang} did not explore the deep near-extremal regime
(\ref{Eq11}) of the rotating Kerr black holes \cite{AZ}. As a
consequence, the co-rotating DMs (\ref{Eq7}) in the regime
$\mu>\mu_{\text{c}}$ have not been observed in the numerical study
of \cite{Yang}. This simple fact has probably led Yang et. al.
\cite{Yang} to the erroneous conclusion that co-rotating DMs are
restricted to the limited range $0\leq \mu\lesssim \mu_{\text{c}}$.

\section{Summary}

It is well-known that rapidly-rotating (near-extremal) black holes
are characterized by {\it two} qualitatively distinct sets of
quasinormal resonances: (1) Zero-damping modes (ZDMs), which are
characterized by the asymptotic property \cite{Hod1}
$\Im\omega^{\text{ZDM}}(\tau\to 0)\to 0$, and (2) Damped modes
(DMs), which are characterized by the asymptotic property
$\Im\omega^{\text{DM}}(\tau\to 0)\to {\text{finite values}}$.

In this Comment we have refuted the claim made in Ref. \cite{Yang}
that co-rotating DMs of near-extremal black holes are restricted to
the limited range $0\leq \mu\lesssim \mu_{\text{c}}$ [see Eqs.
(\ref{Eq2}) and (\ref{Eq4})]. In particular, we have pointed out
that the analytical expression (\ref{Eq7}), originally derived in
\cite{Det}, describes DMs in the $\mu>\mu_{\text{c}}$ regime
\cite{Notemu}, the regime which was claimed in \cite{Yang} not to
contain DMs.

Most importantly, we have emphasized the fact that the analytical
expression (\ref{Eq7}) for the black-hole DMs is not expected to be
valid outside the deep near-extremal regime (\ref{Eq11}) of
rapidly-rotating black holes.

Finally, it is worth emphasizing that rapidly-rotating black holes
in the regime (\ref{Eq11}) are probably of no astrophysical
relevance \cite{Thor}. However, these near-extremal black holes are
very important from the point of view of quantum field theory. In
particular, these black holes play a key role in the conjectured
relation between the quantum states of near-extremal black holes and
the corresponding quantum states of a two-dimensional conformal
field theory \cite{Bar,Gui,Stro,Notesee}.

\bigskip
\noindent {\bf ACKNOWLEDGMENTS}
\bigskip

This research is supported by the Carmel Science Foundation. I would
like to thank Aaron Zimmerman for interesting correspondence. I
would also like to thank Yael Oren, Arbel M. Ongo, Ayelet B. Lata,
and Alona B. Tea for stimulating discussions.


\begin{thebibliography}{99}

\bibitem{Detw} S. L. Detweiler, in Sources of Gravitational Radiation,
edited by L. Smarr (Cambridge University Press, Cambridge, England,
1979).

\bibitem{Yang} H. Yang, F. Zhang, A. Zimmerman, D. A. Nichols, E. Berti, and Y.
Chen, Phys. Rev. D {\bf 87}, 041502(R) (2013). See also, H. Yang, A.
Zimmerman, A. Zenginoglu, F. Zhang, E. Berti, and Y. Chen, Phys.
Rev. D {\bf 88}, 044047 (2013).

\bibitem{Hod1} For these unique black-hole resonances, see:
S. Hod, Phys. Rev. D {\bf 75}, 064013 (2007) [arXiv:gr-qc/0611004];
S. Hod, Class. Quant. Grav. {\bf 24}, 4235 (2007) [arXiv:0705.2306];
S. Hod, Phys. Rev. D {\bf 78}, 084035 (2008) [arXiv:0811.3806]; S.
Hod, Phys. Rev. D {\bf 80}, 064004 (2009) [arXiv:0909.0314].

\bibitem{Notetem} The Bekenstein-Hawking temperatures of Kerr black
holes are given by the relation $T_{\text{BH}}=\tau/8\pi M$. We
shall use units in which $G=c=\hbar=1$.

\bibitem{Hod2} S. Hod, Phys. Lett. B {\bf 715}, 348 (2012) [arXiv:1207.5282].

\bibitem{Delta} The parameter $\delta^2$ is closely related to the
angular-eigenvalue of the angular Teukolsky equation, see
\cite{Teuk} for details [see, in particular, equations (2.7) and
(6.3) of \cite{Teuk}].

\bibitem{Teuk} S. A. Teukolsky and W. H. Press, Astrophys. J. {\bf 193}, 443
(1974).

\bibitem{Det} S. Detweiler, Astrophys. J. {\bf 239}, 292 (1980).

\bibitem{Notemu} That is, as shown in \cite{Det}, the expression (\ref{Eq7}) is valid for co-rotating modes
with real $\delta$ eigenvalues.

\bibitem{Notefal} The failure in \cite{Yang} to observe these resonances
numerically is probably the reason behind the erroneous claim [see
Eq. (\ref{Eq4})] made in \cite{Yang}.

\bibitem{Notesaf} In order to be on the safe side, we have added an
extra order of magnitude to the inequality (\ref{Eq10}).

\bibitem{AZ} A. Zimmerman (private communication) has kindly updated me that he has not detected in
his numerical studies DMs in the regime $\mu>\mu_{\text{c}}$ for a
rapidly-rotating Kerr black hole with ${{a}/{M}}=1-10^{-9}$.

\bibitem{Thor} K. S. Thorne, Astrophys. J. {\bf 191}, 507 (1974).

\bibitem{Bar} J. M. Bardeen and G. T. Horowitz, Phys. Rev. D {\bf 60}, 104030
(1999).

\bibitem{Gui} M. Guica, T. Hartman, W. Song, and A. Strominger, Phys. Rev. D {\bf 80}, 124008 (2009).

\bibitem{Stro} A. Strominger and C. Vafa, Phys. Lett. B {\bf 379}, 99
(1996).

\bibitem{Notesee} For the physical relevance of these near-extremal black holes to the conjectured
universal relaxation bound, see \cite{Hod1} and also: A. Gruzinov,
arXiv:gr-qc/0705.1725; A. Pesci, Class. Quantum Grav. {\bf 24}, 6219
(2007); S. Hod, Phys. Lett. B {\bf 666} 483 (2008)
[arXiv:0810.5419].
\end{thebibliography}
\end{document}